\documentclass[aps,prd,showpacs,twocolumn,superscriptaddress,nofootinbib,preprintnumbers]{revtex4-1} 

\makeatletter
\def\p@subsection{}
\makeatother

\usepackage{graphicx,amsmath,amssymb,lipsum,bm}

\usepackage[usenames,dvipsnames]{xcolor}
\definecolor{xlinkcolor}{rgb}{0.7752941176470588, 0.22078431372549023, 0.2262745098039215}

\usepackage[colorlinks=true,citecolor=xlinkcolor,linkcolor=xlinkcolor,urlcolor=xlinkcolor, backref=false,pdfborder={0 0 0}]{hyperref}

\usepackage[normalem]{ulem}
\usepackage[utf8]{inputenc} 
\usepackage{mathtools}
\usepackage{aas_macros}
\usepackage{float}

\usepackage{multirow}
\usepackage{ulem}
\usepackage{diagbox}

\usepackage[dvipsnames]{xcolor}
\usepackage{mathrsfs}

\newcommand{\be}{\begin{equation}}
\newcommand{\ee}{\end{equation}}
\newcommand{\beqa}{\begin{eqnarray}}
\newcommand{\eeqa}{\end{eqnarray}}

\newcommand{\bseq}{\begin{subequations}}
\newcommand{\eseq}{\end{subequations}}

\renewcommand{\ln}{\mathop{\rm ln}\nolimits}

\newcommand{\ld}{\Lambda{\rm CDM}}

\newcommand{\kmsMpc}{{\rm km\,s^{-1}\,Mpc^{-1}}}

\def\gsim{\raise0.3ex\hbox{$\;>$\kern-0.75em\raise-1.1ex\hbox{$\sim\;$}}}
\def\lsim{\raise0.3ex\hbox{$\;<$\kern-0.75em\raise-1.1ex\hbox{$\sim\;$}}}

\def\beqn#1{\begin{equation}\label{#1}}
\def\eeqn{\end{equation}}

\def\beqa#1{\begin{eqnarray}\label{#1}}
\def\eeqa{\end{eqnarray}}

\begin{document}

\title{Late-time reconstruction of non-minimally coupled gravity with a smoothness prior}

\author{Gen Ye}
\email{gen.ye@unige.ch}
\affiliation{Universit\'e de Gen\`eve, D\'epartement de Physique Th\'eorique and Centre for Astroparticle Physics, 24 quai Ernest-Ansermet, CH-1211 Gen\`eve 4, Switzerland}

\author{Anton Chudaykin}
\email{anton.chudaykin@unige.ch}
\affiliation{Universit\'e de Gen\`eve, D\'epartement de Physique Th\'eorique and Centre for Astroparticle Physics, 24 quai Ernest-Ansermet, CH-1211 Gen\`eve 4, Switzerland}

\author{Camille Bonvin}
\email{camille.bonvin@unige.ch}
\affiliation{Universit\'e de Gen\`eve, D\'epartement de Physique Th\'eorique and Centre for Astroparticle Physics, 24 quai Ernest-Ansermet, CH-1211 Gen\`eve 4, Switzerland}

\author{Martin Kunz}
\email{martin.kunz@unige.ch}
\affiliation{Universit\'e de Gen\`eve, D\'epartement de Physique Th\'eorique and Centre for Astroparticle Physics, 24 quai Ernest-Ansermet, CH-1211 Gen\`eve 4, Switzerland}

\begin{abstract}
We present a non-parametric, model-independent reconstruction of the cosmological background and perturbation dynamics in non-minimally coupled theories of gravity. Within the Effective Field Theory (EFT) of dark energy framework, we reconstruct the time-dependent cosmological constant, $\Lambda(t)$, and the non-minimal coupling function, $\Omega(t)$, from cosmological data. 
To ensure stability, we apply a correlated smoothness prior that restricts the reconstruction to the space of sufficiently smooth functions.
Using CMB, DESI BAO, Type Ia supernovae, CMB–ISW lensing cross-correlations, and large-scale 3×2pt DES Year 3 data, we find a $2.8\sigma$ hint for a non-minimal coupling. For the dark energy equation of state, our results indicate a preference for the existence of crossing of the phantom divide, $w_{\rm DE}=-1$, at $z<0.8$. The non-minimal coupling effect stabilizes dark energy perturbations, providing a viable physical interpretation of the phantom crossing scenario. Our work paves the way for model-agnostic searches for signatures of modified gravity in cosmological data.

\end{abstract}

\maketitle

\section{Introduction}

Dark energy, which drives the observed accelerated expansion of the Universe, remains one of the central open problems in fundamental physics. An important question is whether this phenomenon is due to a constant vacuum energy or is induced by a dynamical component that evolves over cosmic time. Recent measurements from the Dark Energy Spectroscopic Instrument (DESI), in combination with other datasets, indicate a preference for a time-varying dark energy component~\cite{DESI:2024mwx,DESI:2025zgx}.
Interestingly, the favored scenarios involve an evolution of the dark energy equation of state that crosses the so-called phantom divide, defined by $p_{\rm DE} = -\rho_{\rm DE}$.
Realizing such behavior in most simple single-field theoretical models leads to instabilities~\cite{Vikman:2004dc,Carroll:2003st,Hu:2004kh,Creminelli:2008wc}, but it can arise naturally 
in modified gravity theories \footnote{An alternative possibility is to introduce multiple fields, e.g.~\cite{Feng:2004ad,Wei:2005nw,Caldwell:2005ai,Kunz:2006wc,Cai:2007zv,Qiu:2025oop}},
for instance in models where gravity is non-minimally coupled to matter. This has motivated a number of studies investigating the DESI results within the broader context of scalar-tensor theories, see e.g.~\cite{Chudaykin:2024gol,Ye:2024ywg,Ishak:2024jhs,Pan:2025psn,Chudaykin:2025gdn,Wolf:2024stt,Wolf:2025acj,Wolf:2025jed,Wang:2025znm,Wang:2026wrk,Yao:2025wlx,Ye:2025pem,Ye:2024zpk}.

The effective field theory (EFT) of dark energy approach~\cite{Creminelli:2008wc,Gubitosi:2012hu} provides a powerful and unified framework to describe dark energy and modified gravity models characterized by a single scalar degree of freedom. In this formalism, the gravitational action is constructed as a sum of fixed operators that respect the symmetries of the cosmological background in unitary gauge. The relevance of each operator is controlled by the EFT functions which depend only  on time. 
These time-dependent functions encode the cosmic expansion history and the dynamics of perturbations on large scales, offering a model-agnostic description of modified gravity effects.
In this work, we focus on the evolution of linear perturbations and expand the EFT action up to quadratic order in the metric perturbations.

Within the EFT framework, the EFT functions are arbitrary functions of time. However, the current astrophysical data are not sufficiently constraining to provide informative bounds on general time-dependent functions.~\footnote{It is possible, however, to obtain informative constraints on any single EFT function separately when the remaining EFT functions are fixed, see~\cite{Ye:2024ywg}.} 
For this reason, a common strategy is to assume specific parametric forms for their time dependence~\cite{Ishak:2024jhs}.
However, this practice limits the model-independence of the EFT approach. 
Importantly, the EFT functions are expected to be smooth, as they appear in a controlled EFT expansion.
A natural solution, which respects model-independence, is therefore to restrict the reconstruction to the space of smooth functions by using an appropriate smoothness kernel. 

In a Bayesian framework, a natural approach is to specify a prior on the behavior of the EFT functions that reflects the expected smoothness based on theoretical considerations as well as information from previous analyses~\cite{Espejo:2018hxa,Pogosian:2021mcs,Raveri:2021dbu}.

In this work, we adopt a correlated prior approach following~\cite{Crittenden:2011aa,Raveri:2019mxg}.
In this formalism, the reconstructed function is presented as a Gaussian random variable with a proposed correlation function.
The correlation function correlates neighboring bins at separations smaller than a specified correlation length, while allowing for smooth behavior over larger time intervals.
The correlated prior acts as a frequency-dependent filter: it penalizes high-frequency contributions, while allowing physically relevant low-frequency modes to fit the data without penalty. 
This approach enables a stable inference of the time-dependent EFT functions without assuming specific parametric time dependence~\cite{Raveri:2019mxg}.

The correlated prior approach offers several advantages over standard reconstruction methods.
First, it enables systematic control of overfitting, a generic problem for non-parametric reconstruction approaches.
The smoothness prior effectively penalizes high-frequency variations of the functions, which are typically induced by sample noise in the data.
At the same time, physically relevant low-frequency modes remain largely unaffected by the prior, and are constrained by the data. 
Second, the correlated prior eliminates the dependence on the binning scheme assumed in the reconstruction. 
In contrast, many commonly used methods (e.g., those based on Gaussian processes~\cite{Holsclaw:2010nb,Holsclaw:2010sk}) lack binning independence, which complicates the interpretation of their results. Overall, the correlated prior approach provides a coherent Bayesian framework for efficient non-parametric reconstruction of gravitational theories.

In this work, we present a proof-of-principle demonstration of the smoothness prior approach by considering a specific subclass of scalar–tensor Horndeski theories -- the Generalized Brans–Dicke (GBD) models. This class is characterized by a single scalar degree of freedom non-minimally coupled to gravity, with a general scalar field potential and a canonical kinetic term. Models of this family include several well-known theories of gravity, such as Jordan–Brans–Dicke~\cite{Brans:1961sx}, $f(R)$~\cite{Hu:2007nk}, chameleon~\cite{Khoury:2003rn}, and symmetron~\cite{Hinterbichler:2010es} models. Importantly, the GBD framework can accommodate a stable crossing of the phantom divide, $w_{\rm DE}=-1$, as suggested by results from the DESI collaboration~\cite{DESI:2025zgx} and other independent analyses~\cite{Chudaykin:2024gol,Ye:2024ywg}. Therefore, this class of theories provides a minimal and physically motivated framework to investigate the implications of current cosmological observations.

We present a non-parametric late-time reconstruction of the time-dependent cosmological constant, $\Lambda(t)$, and the non-minimal coupling function, $\Omega(t)$, which fully determine the expansion history and linear perturbations in GBD models.
We derive constraints from the Planck CMB~\cite{Rosenberg:2022sdy}, Planck+ACT CMB lensing~\cite{Carron:2022eyg}, DESI DR2 BAO~\cite{DESI:2025zgx}, Pantheon+ supernovae~\cite{Scolnic:2021amr}, CMB–ISW lensing cross-correlations~\cite{Carron:2022eyg}, and the three large-scale two-point correlation functions (3$\times$2pt) from the DES Year 3 data~\cite{DES:2021wwk}.

Our work improves the previous reconstruction of the EFT functions~\cite{Raveri:2019mxg} in several directions.  
First, we incorporate up-to-date cosmological datasets, including Planck measurements based on the latest PR4 maps and the DESI DR2 BAO. 
Second, we extend the DES Year 3 3$\times$2pt likelihood to general modified gravity scenarios by consistently employing the Weyl potential auto- and cross-spectra.
The DES photometric measurements provide additional information on the growth of perturbations that significantly improve the reconstruction results.  
Third, we adopt a modified implementation of the smoothness prior, which optimizes the choice of the prior parameters.

The remainder of this paper is structured as follows.
In Sec.~\ref{sec:method} we introduce the correlated prior approach, describe our analysis pipeline and the datasets employed in the analysis.
In Sec.~\ref{sec:res} we present the main cosmological results. We conclude in Sec.~\ref{sec:conc}. Appendix~\ref{app:prior} explores the sensitivity of the reconstruction results to the choice of prior parameters.
In Appendix~\ref{sec:fs8}, we discuss the compatibility of the reconstruction results with redshift-space distortion measurements. 
Appendix~\ref{app:lcdm} presents constraints on $\ld$ parameters.

\section{Data and Methodology}\label{sec:method}

In order to parameterize evolving dark energy and its modified gravity effect in a theoretically consistent way, we base the reconstruction on the EFT of dark energy action, up to linear perturbation order
\cite{Bloomfield:2012ff,Gubitosi:2012hu}
\begin{equation} 
\begin{split}
    S &= \int d^4x\,\sqrt{-g}\Biggl\{ \frac{M_p^2}{2}\Bigl[ 1 + \Omega(t) \Bigr] R + \Lambda(t) - c(t)\,a^2\,\delta g^{00} \\
    &+ \frac{M_p^2 H_0^2}{2}\,\gamma_1(t) \left( a^2\,\delta g^{00} \right)^2
    -\frac{M_p^2 H_0}{2}\,\gamma_2(t)\,a^2\,\delta g^{00}\,\delta K^\mu{}_\mu\\[1mm]
    &-\frac{M_p^2}{2}\,\gamma_3(t)\Biggl[
    \left( \delta K^\mu{}_\mu \right)^2 - \delta K^\mu{}_\nu\,\delta K^\nu{}_\mu
    -\frac{a^2}{2}\,\delta g^{00}\,\delta R
    \Biggr]
    \Biggr\} \\
    &+ S_m\left[g_{\mu\nu},\chi_m\right].\label{eq:eft}
\end{split}
\end{equation}
 In contrast to the phenomenological descriptions, such as an effective fluid in the parameterized post-Friedmann framework, embedding into a consistent theory \eqref{eq:eft} allows us to discuss theoretical stability of the dark energy / modified gravity theory. The EFT functions $\{\Omega,\Lambda,\gamma_{1,2,3}\}$ are functions of time only and $c$ is not independent once $\Omega$  and $\Lambda$ are known. This parametrization has the advantage that $\{\gamma_{1,2,3}\}$ impacts only perturbations while $\{\Omega,\Lambda\}$ affects both background and perturbations. Another desirable feature is that non-minimal coupling of gravity is uniquely determined by the EFT function $\Omega$. Given the discussion in the previous section, here we are interested in reconstructing the time-dependent dark energy evolution and the associated non-minimal coupling effect. To consistently reconstruct both effects the minimal starting point in terms of Eq.\ \eqref{eq:eft} is reconstructing $\{\Omega,\Lambda\}$ as functions of time while fixing $\{\gamma_{1,2,3}\}$ to zero. Theoretically this corresponds to generalized Jordan-Brans-Dicke theories, i.e.
\begin{equation}\label{eq:jbd}
    \mathcal{S}=\int dx^4\sqrt{-g}\left[ F(\phi)R+X-V(\phi)\right] + \mathcal{S}_m[g_{\mu\nu},\chi_m]\,,
\end{equation}
which is a subclass of the Horndeski theory with luminal gravitational waves \cite{Creminelli:2017sry,Ezquiaga:2017ekz}. Here $X\equiv-\frac{1}{2}\partial_\mu\phi\partial_\nu\phi g^{\mu\nu}$ is the canonical kinetic term. A thorough consideration of all $c_T^2=1$ Horndeski theories requires also varying the non-canonical kineticity $\gamma_1$ and the kinetic braiding $\gamma_2$, corresponding to $G_2(X,\phi)$ and $G_3(X,\phi)\Box\phi$. In this paper we consider only reconstruction of $\Omega$ and $\Delta\Lambda/\Lambda_0\equiv(\Lambda-\Lambda_0)/\Lambda_0$, with $\Lambda_0 = -\rho_{\rm DE}(a=1)$ being the dark energy density today, and assume $\gamma_{1,2,3}=0$. The more complete case will be studied in an upcoming paper.

We consider only viable gravitational theories and impose the no-ghost and no-gradient stability conditions. 

\subsection{The correlated prior}\label{sec:prior}
We perform a non-parametric reconstruction of the time-dependent EFT functions. 
To ensure a stable reconstruction, we employ a correlated prior approach. 
This method imposes a prior on the evolution of the reconstructed functions that effectively enforces the smoothness of the reconstruction. 

We formulate the reconstruction method for a general time-dependent function, $f(t)$.
For simplicity, we express the time dependence of the function in terms of the scale factor, $f(a)$. We then assume that $f(a)$ can be described as a Gaussian random field. In this case, the prior distribution of $f(a)$ is fully characterized by a correlation function that describes fluctuations around a mean or fiducial model, $\bar f(a)$,
\be\label{eq:corr}
\xi(|a - a'|) \equiv \langle [f(a) - \bar f(a)] [f(a') - \bar f(a')] \rangle\,,
\ee
where $\langle..\rangle$ denotes an average over the ensemble of possible realizations of the random function. Here, we assume that the correlation function is translation invariant, depending only on the difference in scale factor $|a-a'|$. The amplitude of the correlation function determines the strength of the prior, while its shape specifies how rapidly correlations decay with distance (in our case, with time separations).
To specify the correlated prior, one must choose the parameterization of the correlation function. 

In this work, we adopt the Crittenden-Pogosian-Zhang (CPZ) prior~\cite{Crittenden:2005wj,Crittenden:2011aa},
\be\label{eq:CPL}
\xi(\delta a)=\frac{\xi_0}{1+(\delta a/a_c)^2}\,,
\ee
where $a_c$ represents the characteristic smoothing scale (correlation length), and $\xi_0$ is a normalizing factor that relates to the expected variance of $f(a)$.
The correlation falls off with distance, so the prior mainly constrains the high frequency modes, while leaving slowly varying modes largely unaffected.
The shape of the correlation function determines the spectrum of eigenvalues of the corresponding prior matrix. The CPZ prior leads to eigenvalues that are approximately uniformly distributed in logarithmic space, with the spacing (the steepness of the spectrum) directly controlled by $a_c$.
Alternative choices for the correlation function (e.g. exponential or power-law forms) result in a spectrum with a mode-dependent slope~\cite{Crittenden:2011aa}. 
Following~\cite{Raveri:2019mxg}, we adopt the CPZ prior for simplicity.

For numerical implementation, we represent the reconstructed function on a fixed time grid. To illustrate the method, we choose a uniform binning in the scale factor $a$, with bin width $\Delta=a_{i+1}-a_i$. To capture non-trivial behavior of $f(a)$, we define the bin-averaged function, 
\be\label{eq:fbin}
f_i = \frac{1}{\Delta} \int_{a_i}^{a_i+\Delta} da\, f(a) \, .
\ee
We then average the correlation function within each pair of bins, yielding the discretized prior covariance, 
\be\label{eq:cov}
C_{ij} =  \frac{1}{\Delta^2} \int_{a_i}^{a_i+\Delta} da \int_{a_j}^{a_j+\Delta} da'\, \xi\!\left(\lvert a - a' \rvert\right) \, .
\ee
Finally, we compute the prior contribution to the total posterior $\chi^2=\chi^2_{\rm data}+\chi^2_{\rm prior}$,
\be\label{eq:chi2}
\chi_{\rm prior}^2=-\frac{1}{2}\sum_{i,j}(f_i-\bar{f}_i)C_{ij}^{-1}(f_j-\bar{f}_j)\,.
\ee
In the case of multiple functions, the total prior contribution is obtained by summing over all functions, $f$.

Physically, the correlated prior effectively binds together neighboring bins at separations smaller than the correlation length $a_c$. 
This enforces smoothness in the reconstruction by disfavoring rapid variations, while allowing functions that vary smoothly over larger time intervals.

The last ingredient of the correlation prior is the fiducial model, $\bar f_i$. Formally, the fiducial model represents the most probable configuration that maximizes the prior probability. Following~\cite{Raveri:2019mxg}, we implement a floating fiducial model obtained using a Gaussian smoothing kernel with a variance matched to the correlation length of the prior, 
\be
\bar{f}_i=\frac{\sum_j f_je^{-(a_i-a_j)^2/2a_c^2}}{\sum_j e^{-(a_i-a_j)^2/2a_c^2}}\, .
\ee
This introduces additional flexibility into the fiducial model while enforcing smooth behavior over large time separations.
As demonstrated in~\cite{Crittenden:2011aa}, fiducial models adopting alternative smoothing schemes produce very similar results. 

The normalization of the prior can be related to the expected average variance of $f(a)$ over the full time interval, defined as 
\be\label{eq:sigma}
\sigma_f^2 \equiv \int_{a_{\min}}^1 \int_{a_{\min}}^1 \frac{da\, da'\, \xi(a - a')}{(1 - a_{\min})^2}
\;\simeq\; \frac{\pi \,\xi_0\, a_c}{1 - a_{\min}} \, ,
\ee 
where $a_{\rm min}$ is the minimal value of the scalar factor used in the late-time reconstruction.
In what follows, we determine the strength of the prior by specifying $\sigma_f$.

Overall, the CPZ prior is specified by two parameters: the correlation length $a_c$ and the average variance of the reconstructed function, $\sigma_f^2$.
The choice of these prior parameters should reflect theoretical expectations and incorporate information from earlier cosmological data that are not explicitly included in the present analysis. 
We detail our selection of prior parameters in the next section.

\subsection{Analysis procedure}\label{sec:pipe}

{\bf Late-time reconstruction:} In this work, we carry out a late-time reconstruction of the time-dependent EFT functions $f=\{\Omega(a),\Delta\Lambda(a)/\Lambda_0\}$, where $\Lambda_0$ is the present-day cosmological constant. The EFT functions are specified by their values on a fixed time grid, which are then interpolated using a quintic spline. The main grid consists of ten equally spaced nodes in the scale factor, 
\be\label{eq:ai}
a_i=\{0.1,0.2,\dots,1\}\quad i=1,\dots,10
\ee

We denote the EFT functions evaluated at these nodes as $\widehat f_i=f(a_i)$~\footnote{To avoid notational clutter, we will refer to the nodes $\Delta\Lambda(a_i)/\Lambda_0$ simply as $\hat{\Lambda}_i$.}.
Since we focus on late-time modifications of gravity, we enforce general relativity in the asymptotic past by fixing $\Omega|_{a=0}=\Delta\Lambda/\Lambda_0|_{a=0}=0$. 
To guarantee a smooth connection between the first node $a_1=0.1$ and the asymptotic behavior at $a_0=0$, we introduce an auxiliary node at $a_{1/2}=0.05$ with its value fixed to $\widehat f_1/2$.
In total, the interpolation is performed over twelve grid points $a_i \in [0,1]$ with $i=0,1/2,1,\dots,10$, among which the last 10 nodes are freely varied in a Monte-Carlo Markov Chain (MCMC) analysis.
The use of a quintic spline interpolation ensures that the reconstruction functions, along with their first and second derivatives entering the dynamical equations, are smooth. 

We now discuss the parameters of the smoothness prior. We adopt a correlation length $a_c=0.3$ for both EFT functions. As shown in~\cite{Raveri:2019mxg}, this choice ensures a stable reconstruction of the late-time behavior of the EFT functions. It corresponds to approximately three node points per correlation length, making the reconstruction robust to the choice of binning~\cite{Raveri:2019mxg}.
This choice allows the EFT functions to have nearly two distinct extrema within the interval $a\in[0.1,1]$, providing sufficient flexibility.~\footnote{It is important to note that the dark energy equation of state, $w_{\rm DE}(a)$, can exhibit more rapid variations, as it involves time derivatives. }
As demonstrated in Appendix~\ref{app:prior}, adopting a less restrictive correlation length, $a_c=0.2$, leads to very similar results.

For the strength of the prior, we adopt the following choices. First, for the non-minimal coupling, previous studies constraining $\Omega(a)$, based on both parametric~\cite{Planck:2018vyg} and non-parametric~\cite{Pan:2025psn} reconstruction techniques, showed that cosmological data impose tight bounds on the non-minimal coupling, with a typical variance being smaller than $0.1$ in the similar time interval. In our analysis, we adopt therefore $\sigma_{\Omega}=0.1$, ensuring that the prior is sufficiently flexible and encompasses the viable range of variations in $\Omega(a)$ informed by the data. Second, for the time-dependent cosmological constant, we adopt $\sigma_\Lambda=0.2$. This choice is motivated by previous non-parametric reconstruction analyses~\cite{Pogosian:2021mcs,Raveri:2019mxg,deBoe:2024gpf,Ye:2024ywg,DESI:2025fii,Berti:2025phi} which report typical variations in the cosmological expansion history of order $\simeq 0.1-0.2$ at late times.~\footnote{Notably, these analyses do not enforce smoothness in the reconstruction, implying that the physically relevant variations in the background history are likely smaller.} 
Increasing the variance to $\sigma_\Lambda=0.4$ does not significantly change the reconstruction results (as validated in Appendix~\ref{app:prior}).
We conclude that our choice of $\sigma_\Lambda$ is conservative.
 
Our choice of the prior normalization differs from that adopted in~\cite{Raveri:2019mxg}. In that work, the overall strength of the prior was fixed by imposing a condition on the eigenvalues of the corresponding prior matrix. In contrast, we determine the normalization factor by relating it to the expected variance of the EFT functions through~\eqref{eq:sigma}, as inferred from previous reconstruction analyses. This selection procedure offers several advantages. First, it eliminates the need for additional MCMC runs, as required in~\cite{Raveri:2019mxg}, which simplifies the analysis. Second, it provides a more direct connection to the evolution of the reconstructed functions, directly linking the amplitude of the prior, $\xi_0$, to the average variance, $\sigma^2$, measured from the data.

{\bf MCMC analysis:} Our parameters and priors are listed in Table~\ref{tab:params}.
\begin{table}[!t]
    \centering
    \begin{tabular}{l@{\hspace{2em}}l@{\hspace{2em}}l}
    \hline\hline
    Parameter & Default  & Prior \\
    \hline
        $H_0$  & --- &   $\mathcal{U}[20, 100]$  \\ 
        $\omega_c$  &  --- &  $\mathcal{U}[0.001, 0.99]$ \\
        $\omega_b$  & --- &   $\mathcal{U}[0.005, 0.1]$ \\
        $\ln(10^{10} A_\mathrm{s})$  &  --- &  $\mathcal{U}[1.61, 3.91]$ \\
        $n_s$   & --- &   $\mathcal{U}[0.8,1.2]$ \\
        $\tau$ & --- &  $\mathcal{U}[0.01, 0.8]$ \\
    %\hline
    $\widehat\Omega_1$, ..., $\widehat\Omega_{10}$ & --- & $\mathcal{U}[-2, 2]$ \\
    $\widehat\Lambda_1$, ..., $\widehat\Lambda_{10}$ & --- & $\mathcal{U}[-2, 2]$ \\
    \hline
        $a_c$ & $0.3$ & --- \\
        $\sigma_\Omega$ & $0.1$ & --- \\
        $\sigma_\Lambda$ & $0.2$ & --- \\
    \hline
    \end{tabular}
    \caption{Parameters with their priors used in our analysis. Here, $\mathcal{U}$ refers to a uniform prior in the given range. The Hubble parameter is quoted in $\kmsMpc$ units.
    The parameters in the upper section are directly sampled in our MCMC chains, while the smoothness prior parameters in the lower section are held fixed. 
    }
    \label{tab:params}
\end{table}
We vary the standard six cosmological parameters ($H_0,\omega_c\equiv \Omega_ch^2, \omega_b \equiv \Omega_bh^2, A_s, n_s,\tau$), together with the EFT functions, $\widehat \Omega_i$ and $\widehat \Lambda_i$, evaluated on the fixed equally spaced grid in the scale factor $a \in [0.1,1]$. 

We use $\mathcal{H}$\texttt{-EFTCAMB} \cite{Hu:2013twa,Raveri:2014cka,Ye:2026qqf}, based on the Einstein-Boltzmann solver \texttt{CAMB} \cite{Lewis:1999bs}, to compute the background and linear perturbation cosmology with dark energy/modified gravity described by Eq.~\eqref{eq:eft}. The MCMC model sampling is performed using the cosmological sampler \texttt{Cobaya} \cite{Torrado:2020dgo, 2019ascl.soft10019T}. 
We adopt a Gelman-Rubin convergence criterion $R-1<0.05$ for all sampled parameters, including the nodes of the reconstructed functions. 

{\bf Datasets:} We consider the following datasets:
\begin{itemize}
    \item \textbf{CMB:} The CamSpec version of Planck PR4 high-$\ell$ TTTEEE \cite{Rosenberg:2022sdy} data; Planck 2018 low-$\ell$ TTEE~\cite{Planck:2019nip} data; CMB lensing reconstruction of Planck PR4~\cite{Carron:2022eyg}.
    \item \textcolor{black}{\textbf{BAO:} The DESI DR2 BAO measurement \cite{DESI:2025zgx,DESI:2025zpo}.}
    \item \textbf{SNIa:} Light curve observations of 1550 type Ia Supernovae (SNIa) compiled in the Pantheon+ sample~\cite{Scolnic:2021amr}, with a single nuisance parameter $M_b$, the absolute magnitude calibration of SNIa.
    \item \textbf{ISW}: The integrated Sachs-Wolfe signal in CMB lensing and temperature cross correlation from Planck PR4~\cite{Carron:2022eyg}.
    \item \textbf{DESY3}: The Dark Energy Survey (DES) 3-year 3x2pt galaxy shear and clustering auto and cross correlation data \cite{DES:2021wwk} \footnote{The original likelihood is available only in \texttt{CosmoSis} \cite{Zuntz:2014csq}. It is ported to \texttt{Cobaya} by wrapping \texttt{CosmoSis} as likelihood and theory modules in \texttt{CosmoSis2Cobaya}, available at \url{https://github.com/JiangJQ2000/cosmosis2cobaya}. The original \texttt{CosmoSis2Cobaya} and \texttt{CosmoSis} likelihood module assume general relativity. We generalize both to modified gravity theories. The codes are publicly available at \url{https://github.com/genye00/cosmosis2cobaya} and \url{https://github.com/genye00/cosmosis_mg_public}.}. Following \cite{DES:2022ccp} we use the same linear scale cut and non-linear alignment  intrinsic alignment model because non-linear correction for general dark energy / modified gravity theory is still unresolved. We also use the Weyl potential auto and cross spectra instead of the matter power spectrum for shear and magnification to be consistent with modified gravity. 
\end{itemize}

CMB, BAO, and SNIa measurements are primarily sensitive to the effects of gravitational theories through their effects on geometric quantities (with the important exception of gravitational lensing and the large-scale excess of power in the TT CMB spectrum induced by the late ISW effect).
In contrast, the ISW and DESY3 datasets directly probe the late-time growth of density perturbations, making them sensitive to clustering properties. 
In the following, we refer to the CMB+BAO+SNIa combination simply as ``Base''.

In the main analysis, we present the results for the two data combinations: Base and Base+ISW+DESY3.

\section{Result}\label{sec:res}

\begin{figure*}
    \centering
    \includegraphics[width=0.9\textwidth]{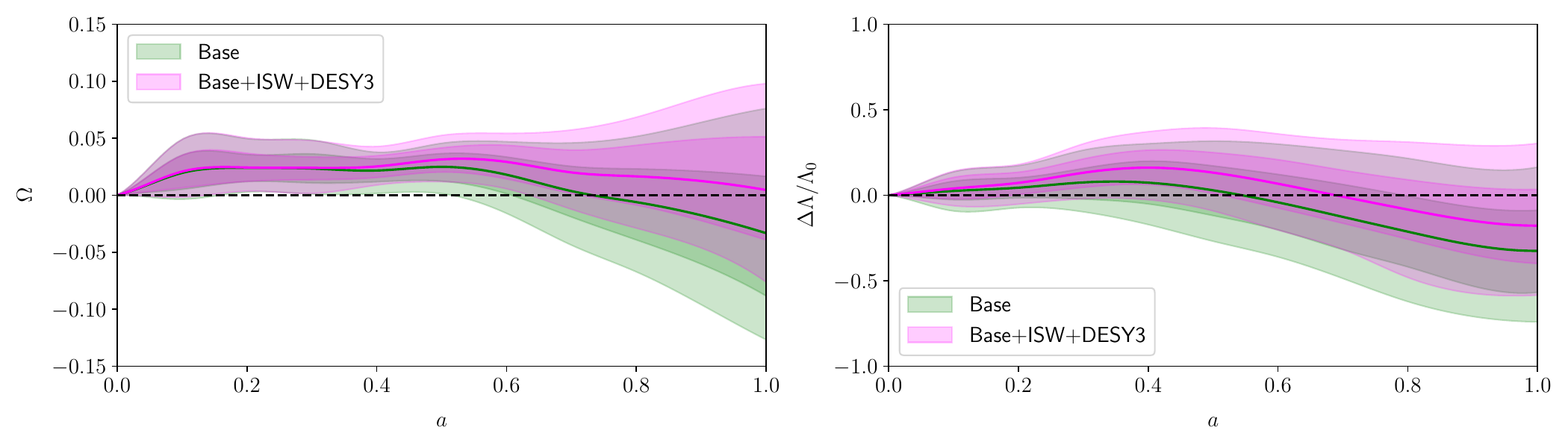}
    \caption{The reconstructed EFT functions $\Omega(a)$ and $\Lambda(a)$ as functions of scale factor. The solid lines plots the mean function. The shaded regions are the $1\sigma$ and $2\sigma$ posterior region of the function value at each scale factor. 
    The Base dataset consists of CMB+BAO+SNIa.
    }
    \label{fig:omg-lmd}
\end{figure*}

Fig.~\ref{fig:omg-lmd} shows the reconstructed EFT functions $\Omega(a)$ and $\Delta\Lambda(a)/\Lambda_0$.~\footnote{We note that GR is assumed in the asymptotic past by fixing $\Omega|_{a=0}=\Delta\Lambda|_{a=0}=0$, such that the reconstructed functions smoothly connect to zero; this can affect the reconstruction results at $a\lesssim 0.1$.} We may immediately conclude that the use of the correlated prior ensures stable reconstruction results by filtering out nonphysical sharp features (cf. Fig.~\ref{fig:priorpars} in Appendix~\ref{app:prior}).

\begin{figure*}
    \centering
    \includegraphics[width=0.9\textwidth]{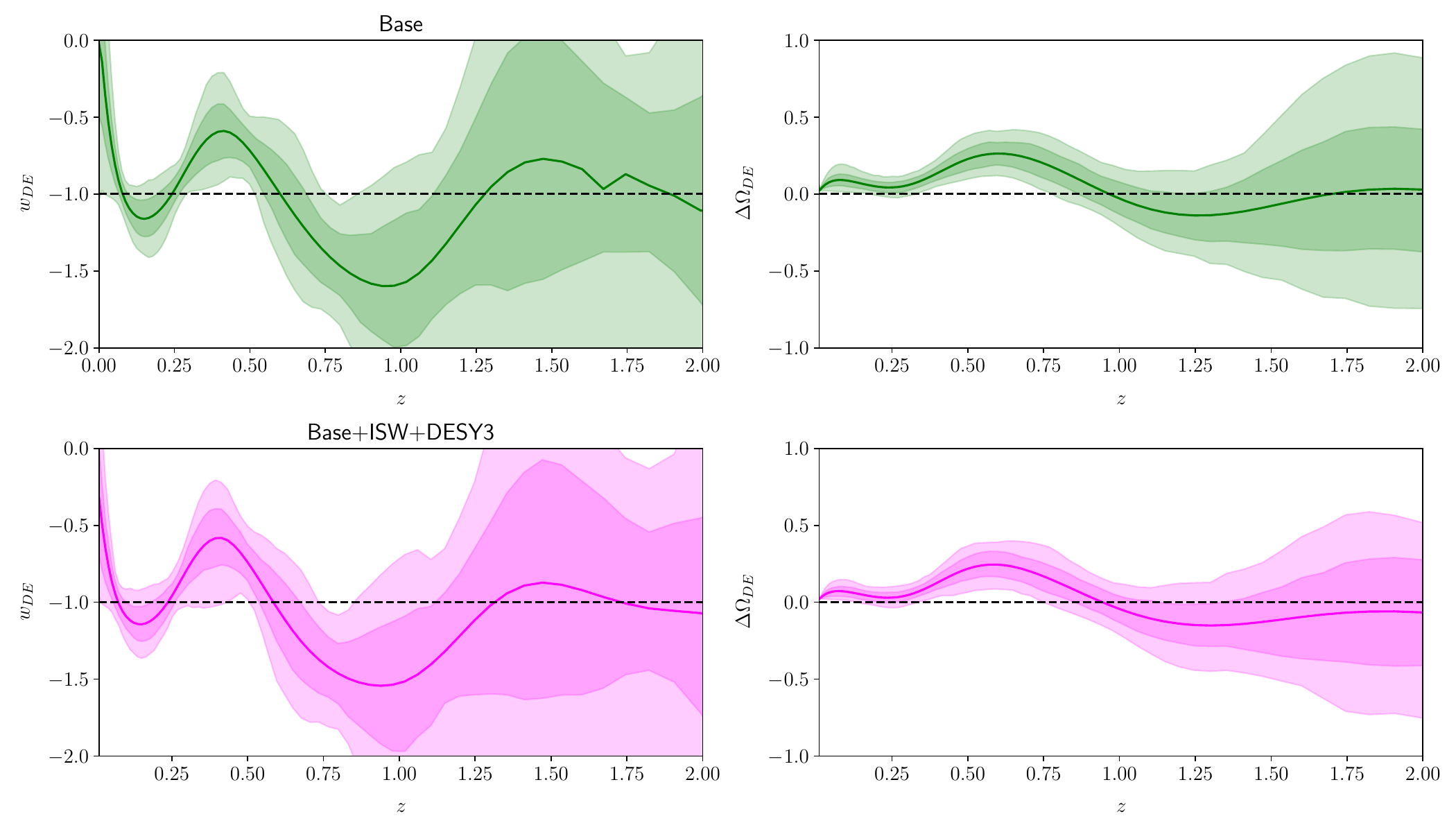}
    \caption{The reconstructed $w_{\rm DE}$ and $\Delta\Omega_X$ as functions of redshift. The solid lines plots the mean function. The shaded regions are the $1\sigma$ and $2\sigma$ posterior region of the function value at each redshift. The gray lines are individual functions from the MCMC sample.
    }
    \label{fig:wde}
\end{figure*}

The results obtained from the Base data show a mild preference for non-minimal coupling at the level of $\Omega\sim0.03$.
Estimating the deviation of $\Omega(a)$ from zero in the region $0.4\lesssim a\lesssim 0.6$ ($0.7\lesssim z\lesssim 1.5$), where the data provide the tightest constraints, we find a $\sim2\sigma$ hint of non-minimal coupling.
This finding is consistent with the results of non-parameteric reconstruction~\cite{Pan:2025psn}, which also observes a bump in $\Omega(a)$; however, the preference for non-minimal coupling is slightly weaker in our case.
The constraints on the time-dependent cosmological constant, $\Lambda(a)$, are weaker than those on $\Omega(a)$.
The reconstructed $\Lambda(a)$ exhibits a peak around $a\simeq 0.4$, which, in the case of $\Omega=0$, implies a phantom crossing behavior: the dark energy equation of state evolves from $w_{\rm DE}<-1$ at early times to $w_{\rm DE}>-1$ at late times.  
Non-minimal coupling is crucial for stabilizing the dynamics of dark energy perturbations and providing a stable phantom crossing scenario. 
For $\Omega\ne0$, the situation becomes more intricate, as the position of the phantom crossing is determined by both $\Omega(a)$ and $\Lambda(a)$.
As discussed later, our results allow for multiple crossings of the phantom divide at $z<0.8$.

The results based on the Base dataset are primarily sensitive to the dynamics of dark energy perturbations through CMB lensing. We checked that removing CMB lensing has only a marginal effect on the reconstruction results. Therefore, the preference for non-minimal coupling is mainly driven by the theoretical stability of the phantom-crossing scenario favored by the CMB+DESI+SN data. DES photometric measurements, as well as the ISW signal from CMB lensing--temperature cross-correlations, provide the missing direct information on the evolution of perturbations, improving the reconstruction results.

Including the DESY3 and ISW data significantly tightens the constraints on both $\Omega(a)$ and $\Lambda(a)$ at $a>0.5$ ($z<1$), corresponding to the redshift range probed by DES Y3.~\footnote{By studying different data combinations, we found that the additional constraining power mainly originates from the DESY3 data, while the effect of ISW is subdominant.} The significance of the $\Omega>0$ bump in the region $0.4\lesssim a\lesssim 0.6$ increases to $2.8\sigma$ after including ISW+DESY3.
The reconstructed $\Lambda(a)$ is shifted toward slightly larger values, making the peak at $a\sim 0.4$ more pronounced when incorporating large-scale structure data.
The results derived from the Base+ISW+DESY3 data are fully consistent with the Base-only results, indicating that the modified perturbation evolution associated with non-minimal coupling is compatible with the DESY3 observations.

\begin{figure*}
    \centering
    \includegraphics[width=1\textwidth]{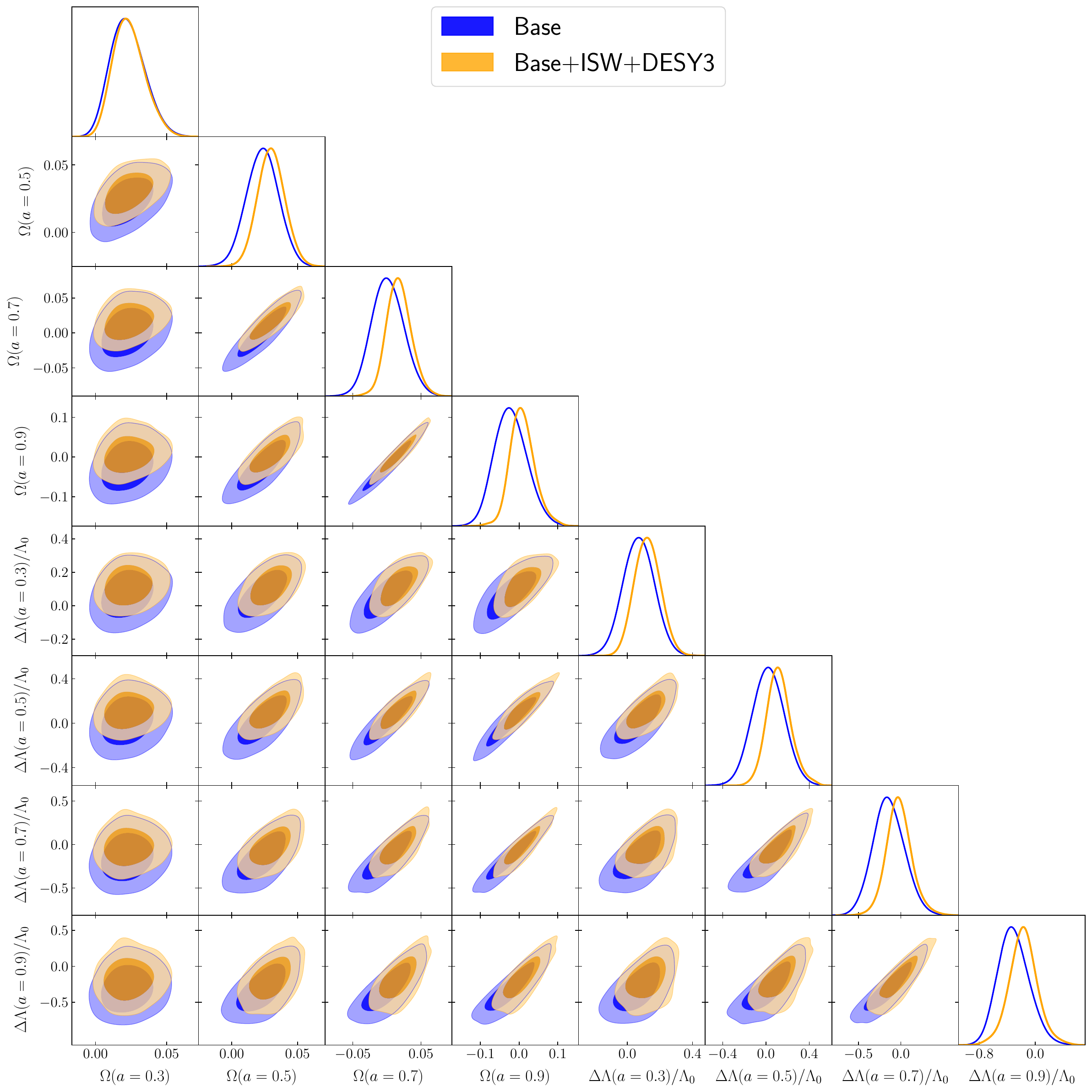}
    \caption{$1\sigma$ and $2\sigma$ posterior distribution of the $\Omega(a)$ and $\Delta\Lambda(a)/\Lambda_0$ reconstruction nodes at $a=0.3,0.5,0.7,0.9$.}
    \label{fig:nodetrig}
\end{figure*}

Fig.\ \ref{fig:nodetrig} shows the one- and two-dimensional marginalized posteriors for the selected reconstructed functions, demonstrating strong correlations between the $\Omega(a_i)$ and $\Delta\Lambda(a_i)/\Lambda_0$ nodes.
As discussed above, the non-minimal coupling is essential for stabilizing the phantom crossing behavior in the cosmological background history, implying strong correlations between the two functions.

Here, we visualize the correlation by plotting the dark energy density $\rho_{\rm DE}$ and equation of state $w_{\rm DE}$, which receive contribution from both $\Omega(a)$ and $\Lambda(a)$. To this end, we define the effective equation of state
\begin{equation}\label{eq:wDE}
    w_{\rm DE}(a)\equiv\frac{ -2\dot{H}-3H^2-P_{\rm{m}} }{ 3H^2-\rho_{\rm{m}} }\, ,
\end{equation}
and dark energy energy fraction
\begin{equation}
    \Omega_X(a)\equiv1-\frac{\rho_{\rm m}}{3M_p^2H^2}\,.
\end{equation}
Fig.~\ref{fig:wde} shows $w_{\rm DE}(a)$ and $\Delta\Omega_X(a)\equiv \Omega_X(a)/\Omega_{\Lambda}(a)-1$ derived from the Base and Base+ISW+DESY3 datasets, where $\Omega_{\Lambda}(a)\equiv \Lambda_0/3M_p^2H^2$. The dark energy equation of state, $w_{\rm DE}$, appears more oscillatory than the reconstructed $\Omega(a)$ and $\Lambda(a)$ shown in Fig.~\ref{fig:omg-lmd}. This occurs because $w_{\rm DE}$ involves the first derivatives of both functions (through $\dot{H}$), allowing for more rapid variations. The inferred $w_{\rm DE}(a)$ confirms a phantom crossing feature near $z\sim0.6$, which manifests as a bump feature in $\Delta\Omega_X(z)$. 
Interestingly, our results also provide mild hints of several phantom crossing episodes at $z<0.6$, although the significance of the corresponding behaviour remains below $2\sigma$.

The shape of $w_{\rm DE}(a)$ is highly consistent with the results of Ref.~\cite{Ye:2024ywg}, obtained using a completely different reconstruction method (i.e.\ reconstructing $w_{\rm DE}(a)$ directly without a smoothness prior for a dark energy fluid). The bump feature in $\Omega_X(a)$ is also consistent with Ref.~\cite{Berti:2025phi}, which directly reconstructed this quantity from the data. Unlike $\Omega(a)$ and $\Lambda(a)$, the inclusion of DESY3 and ISW has only a marginal effect on $w_{\rm DE}(a)$ and $\Omega_X(a)$, because the latter are purely background quantities that are well constrained by the Base data.

In our results, phantom crossing and non-minimal coupling spontaneously emerge from the reconstruction without assuming a parameterization. For the reconstructed GBD theories, the modified gravity parameter modulating the Weyl potential is $\Sigma(a)=(1+\Omega(a))^{-1}$, which is smaller than unity for $\Omega>0$. The modified gravity parameter $\mu$ relevant to clustering also decreases for $\Omega>0$ \cite{Gleyzes:2015rua,Pogosian:2016pwr}. The presence of non-minimal coupling as a bump of $\Omega>0$ near $z=1$ indicates that clustering and lensing is weaker than GR at $z\sim1$ but enhances with time for $z<1$ given the same density perturbation. Our result shows that DESY3, with the linear scale cut, is consistent with this gravity modification and its inclusion improves constraints at $z<1$ thus increases the significance for non-minimal coupling to $2.8\sigma$.

Since our results show a mild deviation from GR, it is important to check that this is not in disagreement with data sets that we have not included in our analysis, in particular redshift-space distortion measurements from spectroscopic surveys. Redshift-space distortions are sensitive to galaxy velocities, which are also affected by a non-minimal coupling. In Appendix~\ref{sec:fs8}, we compare the $f\sigma_8$ measurements from DESI DR1~\cite{DESI:2024jxi} with the best-fit predictions obtained from our analysis. The quantity $f\sigma_8$ is inferred from the data in a relatively model-independent way using the ``ShapeFit'' compression method~\cite{Brieden:2021edu}. We find that the posterior $f\sigma_8$ inferred from our reconstruction agrees very well with the DESI measurements (see Fig.~\ref{fig:fs8}), showing that current velocity measurements are not in tension with the non-minimal coupling inferred from CMB and weak-lensing data.
In a forthcoming paper, we will perform a full-shape analysis of the galaxy clustering statistics, consistently modeling the effects of non-minimally coupled gravity. This will allow us to incorporate the information encoded in the full-shape clustering signal, including redshift-space distortions.

\section{Conclusion}\label{sec:conc}

It is known that non-minimal coupling in gravity can stabilize the phantom-crossing of the dark energy that is hinted at by recent BAO and SNIa observations, which has motivated a number of studies to reconstruct gravity within the EFT of dark energy framework. 
Due to the limited constraining power of cosmological data, EFT-based analyses often impose specific parametric time dependencies for the EFT functions. In this paper, we go beyond this approximation and perform
a parametrization-independent reconstruction of both the background $\Lambda(a)$ and non-minimal coupling function $\Omega(a)$ by applying a smoothness prior that filters out nonphysical sharp features in the reconstructed functions. Our result confirms the $2.8\sigma$ hint for non-minimal coupling even when the background is non-parametrically reconstructed. The existence of non-minimal coupling modifies perturbation evolution. We found that the non-minimal coupling required to stabilize the observed phantom crossing is fully consistent with the DES Y3 galaxy clustering and galaxy lensing observations (3x2pt).

Our analysis presents a proof-of-principle application of the complete late-time reconstruction of the EFT functions using the latest cosmological measurements. We employ a stable framework based on a smoothness prior, which enforces smooth reconstructions and makes the results independent of the binning choice. The method depends on two parameters, the correlation length and the strength of the prior, which can be estimated from analyses of earlier cosmological data. It is straightforward to implement this approach within an MCMC analysis for the reconstruction of general modified gravity theories. The correlated prior reduces the number of independent degrees of freedom and enables a stable inference of smooth but otherwise arbitrary time-dependent functions from cosmological data.

It is important to mention some limitations of the correlated prior approach used in this work. 
First, we assume that the prior distribution is Gaussian, meaning it is fully specified by the correlation function.
Second, we assume that the correlation function is translation invariant and choose the scale factor as the independent variable. 
The reconstruction results may change if a different time variable is used (e.g.\ the redshift $z$). 
Finally, the correlated prior assumes a specific form for the correlation function, the so-called CPZ form. 
Adopting different correlation shapes can affect the reconstruction results~\cite{Crittenden:2011aa}.
Although we expect the reconstruction to be relatively insensitive to such differences provided that the prior parameters are well chosen, it is important to explore the sensitivity of the results to the assumptions discussed above. 

Our analysis can be improved in multiple directions. First, we can explore the broader class of scalar-tensor Horndeski theories. The most general single-field theory of gravity is described by five time-dependent EFT functions, which additionally capture a non-standard kinetic term for the scalar field, kinetic gravity braiding, and deviations of the gravitational wave propagation speed from that of light. Second, the analysis can be improved by incorporating cross-correlations of galaxy shear and galaxy density with the CMB lensing within the 5$\times$2pt DES framework~\cite{DES:2022urg}.
Finally, as already mentioned previously, it would be interesting to include the full-shape galaxy clustering statistics, which have been shown to tighten constraints on modified gravity models~\cite{Ishak:2024jhs}.
We leave these research directions for future exploration.

\vskip 8pt
\acknowledgments
{\small
\begingroup
\hypersetup{hidelinks}
\noindent 
GY, AC, CB and MK acknowledge funding from the Swiss National Science Foundation. The computations were performed at the University of Geneva using the \texttt{Baobab} HPC service. We thank Maria Berti for useful discussions.
\endgroup
}

\appendix

\section{Impact of the smoothness prior parameters}\label{app:prior}

In this Appendix, we explore the robustness of the smoothness prior approach. First, we present the results of a standard reconstruction without assuming the prior. Second, we assess the sensitivity of the reconstruction results to the choice of the prior parameters.

\begin{figure*}
    \centering
    \includegraphics[width=0.8\linewidth]{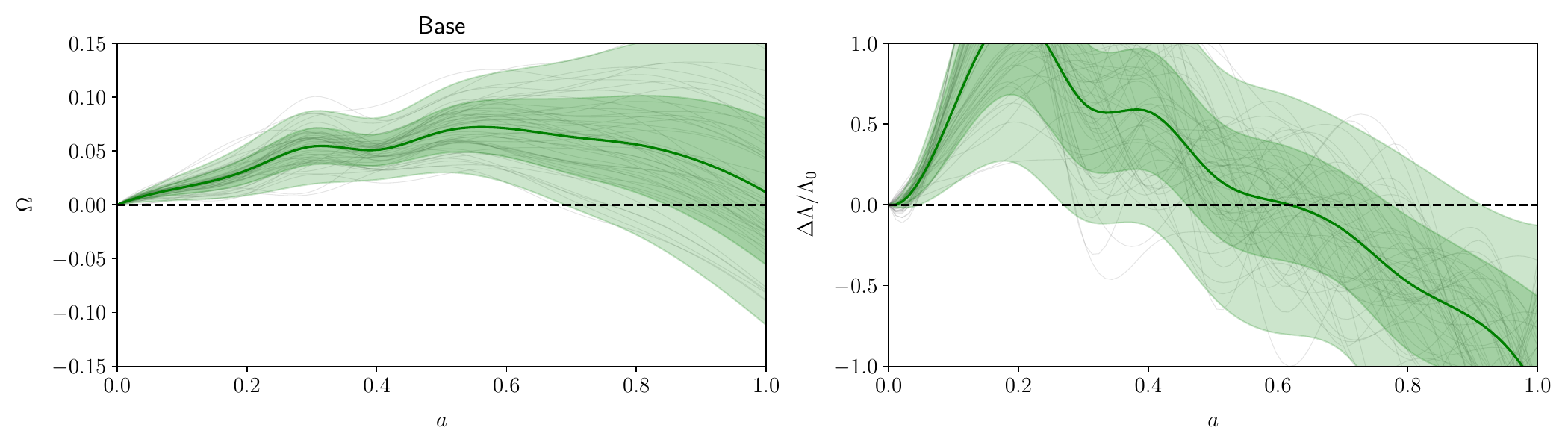}
    \caption{Reconstructed $\Omega(a)$ and $\Delta\Lambda(a)/\Lambda_0$ without the smoothness prior, derived from the Base dataset. The solid lines represent the mean of the reconstructed functions, while the shaded regions indicate the $1\sigma$ and $2\sigma$ confidence intervals. 
    The gray lines show individual realizations randomly selected from the chains. 
    The standard reconstruction on a dense time grid produces significant noise, which biases the $\Delta\Lambda(a)/\Lambda_0$ reconstruction (cf. Fig.~\ref{fig:omg-lmd}). 
    }
    \label{fig:priorpars}
\end{figure*}

\begin{figure*}
    \centering
    \includegraphics[width=0.8\linewidth]{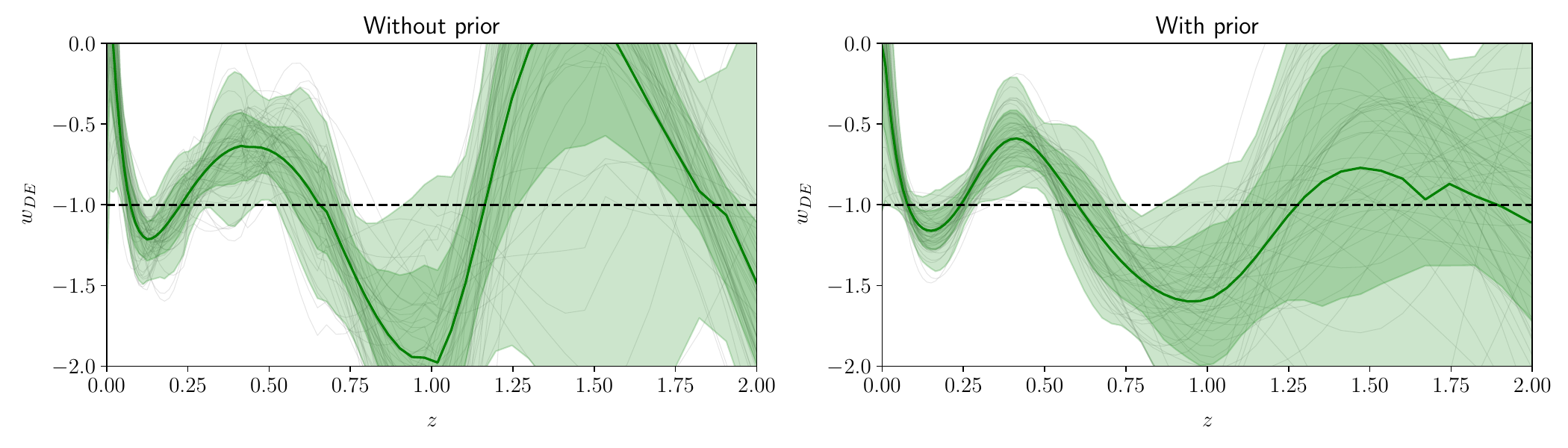}
    \caption{The reconstructed $w_{\rm DE}(z)$ obtained in the analysis without ({\it left panel}) and with ({\it right panel}) the smoothness prior using the Base dataset. The solid lines represent the mean of the reconstructed functions, while the shaded regions indicate the $1\sigma$ and $2\sigma$ confidence intervals. 
    The gray lines show individual realizations randomly selected from the chains. 
    Without the smoothness prior the reconstructed $w_{\rm DE}$ appears more noisy with sharp turns. 
    }
    \label{fig:wde_np}
\end{figure*}

We start by presenting the results obtained without the smoothness prior. 
In this analysis, we vary the cosmological parameters together with the $\widehat\Omega_i$ and $\widehat\Lambda_i$ reconstruction nodes, as specified in Sec.~\ref{sec:pipe}, but without imposing a prior on the evolution of the EFT functions. 
Fig.~\ref{fig:priorpars} shows the reconstructed $\Omega(a)$, $\Delta\Lambda(a)/\Lambda_0$, derived from the Base dataset. 
Fig.~\ref{fig:wde_np} displays the derived $w_{\rm DE}(z)$ obtained using the Base data.

The reconstructed functions obtained without the smoothness prior exhibit nonphysical noise, which significantly biases the recovery of $\Lambda(a)$. 
This noise arises from rapidly varying $\Lambda(a)$ functions that are not well constrained by the data.
The use of a smoothness prior effectively suppresses these solutions, enabling a stable reconstruction of the cosmological expansion history.
The behavior of $\Omega(a)$ is also systematically shifted to higher values compared to the results obtained with the smoothness prior (cf. Fig.~\ref{fig:omg-lmd}), although the corresponding shift is more modest. 
This effect can be attributed to a smoother evolution of $\Omega(a)$ behavior, which is better constrained by the data and theoretical stability.  
Poorly constrained directions in $\Lambda(a)$ without a smoothness prior also propagate into the reconstruction of $w_{\rm DE}(z)$, leading to high-frequency variations in individual realizations that bias the final result.

We conclude that imposing a correlated prior, which enforces a smooth evolution of the reconstructed functions, is essential for obtaining an unbiased and physically meaningful behavior of the time-dependent gravitational dynamics.

\begin{figure*}
    \centering
    \includegraphics[width=0.8\linewidth]{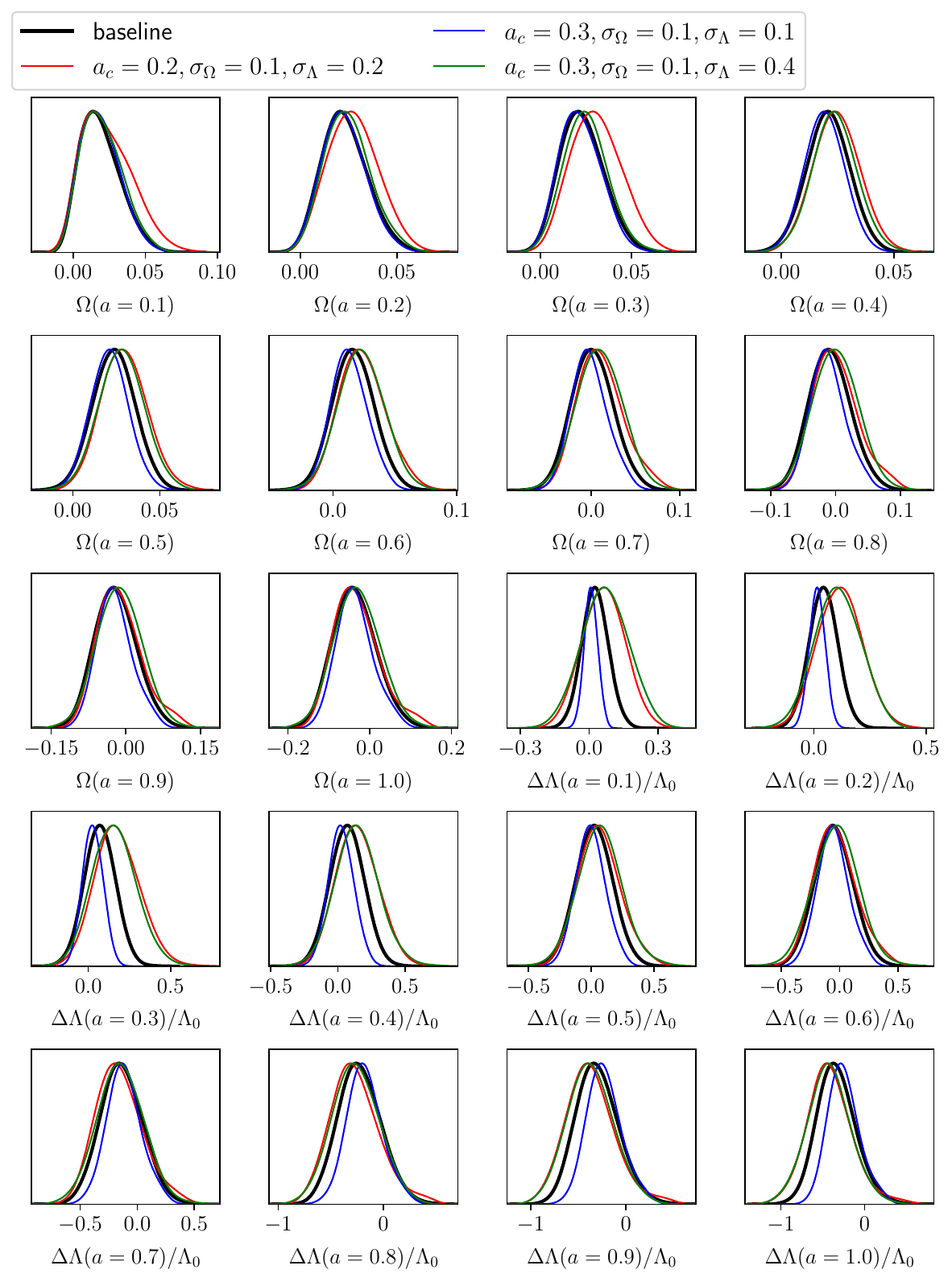}
    \caption{The one-dimensional marginalized posteriors for the $\Omega(a_i)$ and $\Delta\Lambda(a_i)/\Lambda_0$ reconstruction nodes using four different settings of the smoothness prior: (1) the baseline model used in the main analysis (black); (2) a more restrictive correlation length $a_c=0.2$ (red); (3) a two-times smaller variance $\sigma_\Lambda=0.1$ (blue); (4) a two-times larger variance $\sigma_\Lambda=0.4$ (green). The baseline prior settings correspond to $a_c=0.3$, $\sigma_\Omega=0.1$, $\sigma_\Lambda=0.2$. 
    The results are robust to the choice of smoothness prior parameters, though there is the mild sensitivity in the $\Delta\Lambda(a_i)/\Lambda_0$ reconstruction at $a \leq 0.2$. 
    }
    \label{fig:testPr}
\end{figure*}

Next, we explore the sensitivity of the reconstruction results to the choice of prior parameters: a correlation length $a_c$ and an average deviation of the time-dependent cosmological constant $\sigma_\Lambda$.~\footnote{Note that the choice of $\sigma_\Omega$ is already conservative as explained in Sec.~\ref{sec:pipe}, so we do not consider the impact of varying this parameter.}
Fig.~\ref{fig:testPr} shows the one-dimensional posterior distributions of the reconstruction nodes obtained in the Base data analysis for different choices of the prior parameters. 

The reconstruction of the non-minimal coupling function remains robust to variations in $\sigma_\Lambda$. Adopting a smaller correlation length, $a_c=0.2$, introduces additional flexibility in the reconstruction by allowing more rapidly varying functions. This only slightly increases the error-bars on the $\widehat\Omega_i$ nodes with $i=1,2,3$, while the renaming constraints remain largely unchanged.

The $\widehat\Lambda_i$ nodes are also robust with respect to the choice of the prior parameters. The largest deviations arise at the first two reconstruction nodes ($i=1,2$), where the prior lacks sufficient flexibility due to their proximity to the boundary of the interpolation range.
These nodes correspond to early epochs, $z>4$, where the model dynamics is poorly constrained by the data. In this regime, the results are largely prior-dominated, which accounts for the mild sensitivity of the $\Lambda(a)$ reconstruction to the prior parameters. 

Overall, the tests demonstrate that our reconstruction results are robust and that the chosen prior parameters are optimal.

\section{Compatibility with redshift-space distortions\label{sec:fs8}}

\begin{figure}[h]
    \centering
    \includegraphics[width=\linewidth]{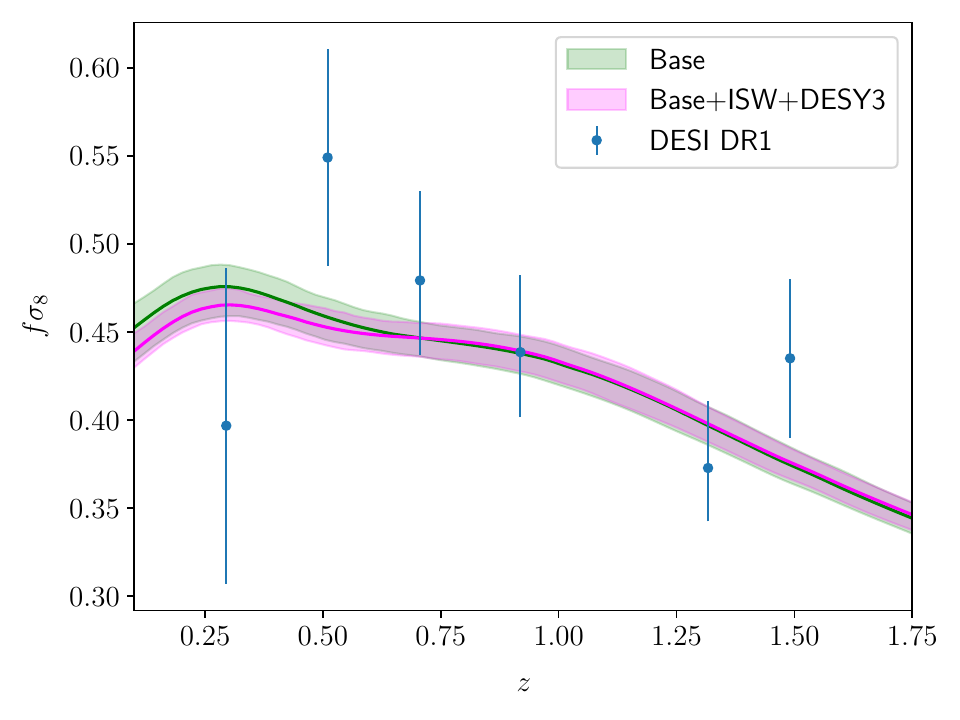}
    \caption{Constraint on the growth parameter $f\sigma_8(z)$ as a function of redshift. The data points are $f\sigma_8$ measured by DESI DR1.}
    \label{fig:fs8}
\end{figure}

Fig.~\ref{fig:fs8} shows the growth parameter $f\sigma_8(z)$ derived from our reconstruction results, together with measured data points by DESI DR1 \cite{DESI:2024jxi}. Compared with Base, inclusion of the DESY3 observation favors less clustering at $z<0.75$. This is consistent with the reconstruction result of $\Omega$ presented in Fig.~\ref{fig:omg-lmd} where DESY3 excludes part of the $\Omega<0$ (stronger gravity) region at low redshift. The $f\sigma_8$ from reconstruction is compatible with current DESI DR1 measurements.

The quantity $f\sigma_8$ has been inferred from the data in a relatively model-independent way, using the ``ShapeFit'' compression method~\cite{Brieden:2021edu}.
This method assumes a fiducial CDM-motivated template for the linear matter power spectrum at high redshift, which is then evolved forward and rescaled by a set of scaling parameters.
These include the Alcock-Paczyński parameters $\alpha_\parallel$ and $\alpha_\perp$, which encode deviations from the fiducial cosmology and regulate the position of the BAO peak, and $f(z)\sigma_8(z)$, which parameterizes the amplitude of peculiar velocities at redshift $z$. In addition, the parameter $m$, modifies the slope of the linear matter power spectrum to marginalize over possible scale-dependent deviations.

In this method $f\sigma_8$ is assumed to be independent of $k$, which is not generally true in modified gravity theories. The dependence in $k$ for most viable EFT of DE models is however expected to be sufficiently weak at the scales we are probing~\cite{Gleyzes:2015rua,Pogosian:2021mcs} -- compared with the current uncertainties on the measurements -- for the method to be adequate for DESI DR1 data.
However, future analyses may require the introduction of additional parameters controlling modifications to the shape of the matter and velocity power spectrum, especially for models with non-trivial scale-dependence.

Importantly, the fixed-template method does not fully utilize the full-shape information present in the linear matter power spectrum that feeds into the observed non-linear galaxy power spectrum.
In a forthcoming paper, we will go beyond this approximation and model the full-shape clustering statistics within the EFT of LSS framework, allowing us to exploit the shape information encoded in the clustering observables.

\section{Constraints on $\Lambda$CDM parameters}\label{app:lcdm}

\begin{figure*}
    \centering
    \includegraphics[width=0.7\textwidth]{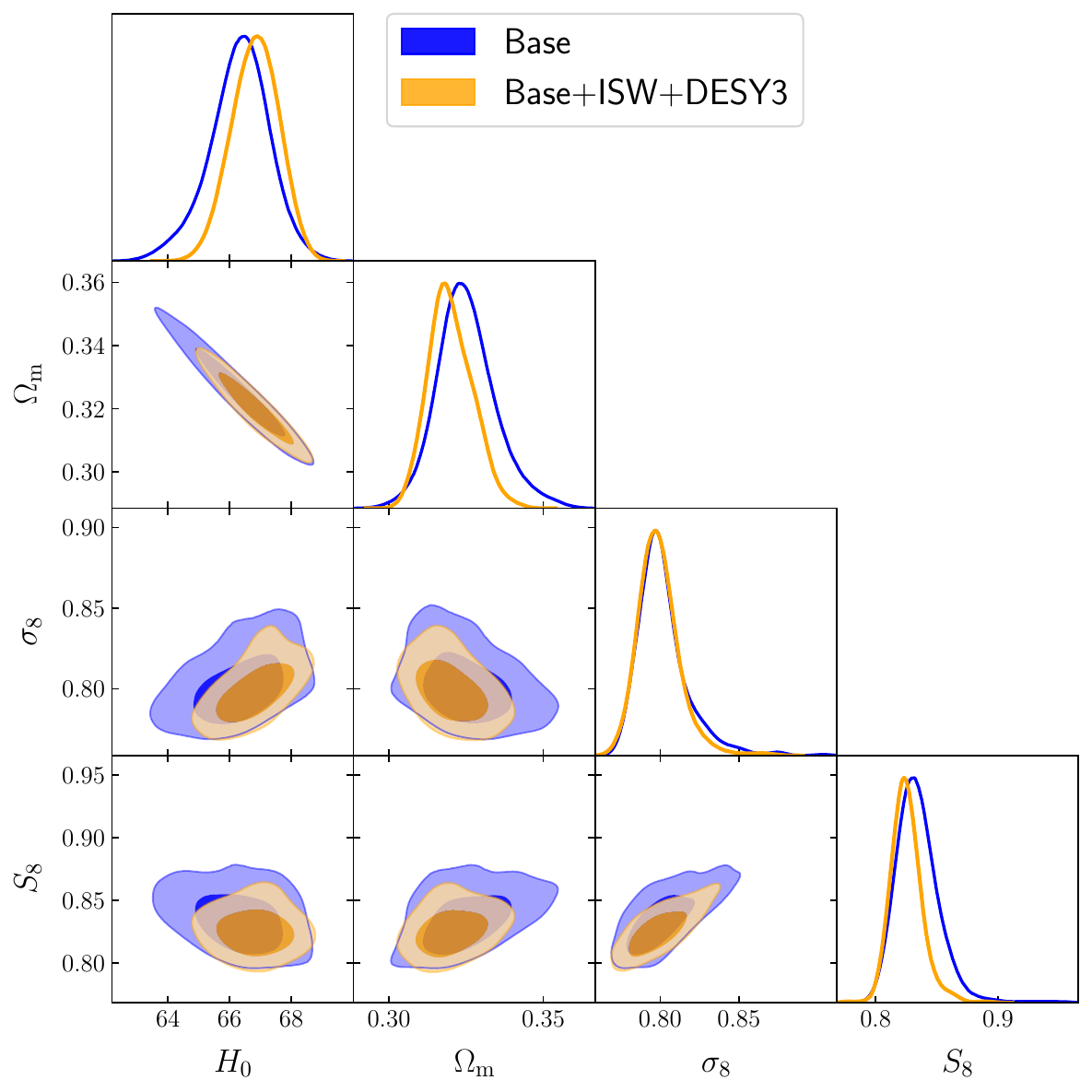}
    \caption{{\bf $\Lambda$CDM parameters:} 1$\sigma$ and 2$\sigma$ posterior distributions for the $\ld$ parameters $H_0$, $\Omega_m$, $\sigma_8$ and $S_8$ obtained in the late-time reconstruction analyses using the Base (blue) and Base+ISW+DESY3 (orange) datasets.}
    \label{fig:cosmo}
\end{figure*}

Here, we present constraints on $\ld$ cosmological parameters obtained in the main analyses of this work. 
Fig.~\ref{fig:cosmo} shows the two-dimensional posterior distributions with one-dimensional marginalized constraints listed in Tab.~\ref{tab:cosmo}.
\begin{table*}[!t]
    \centering
    \begin{tabular}{lcccc}
    \hline
    Data 
    %& $\enspace \ln(10^{10}A_s)\enspace$
    %& $\enspace n_s\enspace$
    & $\enspace H_0\enspace$
    %& $\enspace \omega_b\enspace$
    %& $\enspace \omega_{cdm}\enspace$
    %& $\enspace \tau\enspace$
    & $\enspace \Omega_m\enspace$
    & $\enspace \sigma_8\enspace$
    & $\enspace S_8\enspace$
    \\\hline
    Base
    %& $3.047^{+0.013}_{-0.017}   $
    %& $0.9623\pm 0.0044          $
    & $\enspace 65.3\pm 1.6               \enspace$
    %& $0.02213\pm 0.00016        $
    %& $0.1203\pm 0.0013          $
    %& $0.0563^{+0.0065}_{-0.0087}$
    & $\enspace 0.336^{+0.024}_{-0.019}   \enspace$
    & $\enspace 0.802\pm0.017\enspace$
    & $\enspace 0.834\pm0.019\enspace$
    \\
    Base+ISW+DESY3
    %& $3.040\pm 0.014            $
    %& $0.9621\pm 0.0042          $
    & $66.93^{+0.74}_{-0.95}     $
    %& $0.02210\pm 0.00015        $
    %& $0.1202\pm 0.0012          $
    %& $0.0534\pm 0.0073          $
    & $0.3192^{+0.0086}_{-0.0071}$
    & $0.799\pm0.014$
    & $0.825\pm0.013$
    \\\hline
    \end{tabular}
    \caption{{\bf $\Lambda$CDM parameters:} Mean and 68\% confidence intervals on $\Lambda$CDM cosmological parameters from the main analyses of this work. The constraints on the other cosmological parameters do not change significantly in modified gravity scenarios and therefore are not presented here.
    }
    \label{tab:cosmo}
\end{table*}

Addition of the large-scale structure measurements significantly tightens the parameter constraints. The uncertainties on $H_0$ and $\Omega_m$ are reduced by a factor of two, while the precision of $\sigma_8$ and $S_8$ improves by $20\%$ and $30\%$, respectively, upon adding the ISW+DESY3 data. This improvement is primarily driven by the DESY3 photometric dataset.

The Base+ISW+DESY3 result yields $S_8=0.825\pm0.013$, which is $2.3\sigma$ higher than the DESY3 baseline result in $\Lambda$CDM, $S_{8}=0.776\pm0.017$~\cite{DES:2021wwk}. We attribute this difference to a combination of (a) the modified background and perturbation evolution in our analysis, and (b) the use of the linear scale cuts (which retains 256 out of the 462 data points included in the official DESY3 analysis~\cite{DES:2021wwk}) in our analysis.

\bibliography{reference}

\end{document}